\documentclass[final]{aipproc}

\usepackage{amssymb, amsmath, subfigure}
\usepackage{color}
\usepackage{graphicx}

\def\be{\begin{equation}}
\def\ee{\end{equation}}
\def\bee{\begin{eqnarray}}
\def\eee{\end{eqnarray}}

\def\apj{ApJ}
\def\apjl{ApJL}

\def\nat{Nature}

\def\araa{ARAA}

\def \m {\mathcal{M}}
\def \msun{$\text{M}_{\odot}$} 		%Msun

\newcommand{\ppd}{$\text{P}$-$\dot{\text{P}}$ }

\newcommand{\pdp}{($\dot{P}$) } 

\layoutstyle{8d}

\begin{document}

\title[\textsc{New Constraints on Neutron Stars}]{\textsc{Reassessing The Fundamentals \linebreak \normalsize{New Constraints on the \linebreak Evolution, Ages and Masses of Neutron Stars}}}



\author{B\"ulent K{\i}z{\i}ltan\footnote{Current address: Harvard-Smithsonian Center for Astrophysics, MS-51, Cambridge, MA, 02138-1516; email: bkiziltan@cfa.harvard.edu}}
{address={Harvard-Smithsonian Center for Astrophysics, Cambridge, MA, 02138}, address={Department of Astronomy \& Astrophysics, University of California, Santa Cruz, CA, 95064}}

\begin{abstract}

The ages and masses of neutron stars (NSs) are two fundamental threads that make pulsars accessible to other sub-disciplines of astronomy and physics. A realistic and accurate determination of these two derived parameters play an important role in understanding of advanced stages of stellar evolution and the physics that govern relevant processes. Here I summarize new constraints on the ages and masses of NSs with an evolutionary perspective. I show that the observed \ppd demographics is more diverse than what is theoretically predicted for the standard evolutionary channel. In particular, standard recycling followed by dipole spin-down fails to reproduce the population of millisecond pulsars with higher magnetic fields (B$>4\times 10^{8}$G) at rates deduced from observations. A proper inclusion of constraints arising from binary evolution and mass accretion offers a more realistic insight into the age distribution. By analytically implementing these constraints, I propose a ``modified'' spin-down age ($\widetilde{\tau}$) for millisecond pulsars that gives estimates closer to the true age. Finally,  I independently analyze the peak, skewness and cutoff values of the underlying mass distribution from a comprehensive list of radio pulsars for which secure mass measurements are available. The inferred mass distribution shows clear peaks at 1.35 \msun\,and 1.50 \msun\,for NSs in double neutron star (DNS) and neutron star-white dwarf (NS-WD) systems respectively. I find a mass cutoff at 2 \msun\,for NSs with WD companions, which establishes a firm lower bound for the maximum mass of NSs.
\end{abstract}

\maketitle

\section{\label{sec:evol} Evolution}

MSPs are spun-up second generation neutron stars that have accreted mass and angular momentum from their evolved companion during the LMXB phase \citep{Alpar:82, Radhakrishnan:82}. For these pulsars, I parametrize the transitional phase after the cessation of active accretion and the subsequent evolution. \footnote{For details see \cite{Kiziltan:09, Kiziltan:10, Kiziltan:10g}}$^{,}$\footnote{http://www.kiziltan.org/research.html: Full resolution color figures and movies are available at this URL.}

\subsection{Parametrizing millisecond pulsar evolution}
The equilibrium period distribution ($D$) of millisecond X-ray pulsars (MSXPs) at the end of the accretion phase will be tightly constrained by the Keplerian orbital period at their Alfv\`en radius (Figure 1, cyan line). The geometry combined with the accretion rate ($\dot{M}$) experienced during the LMXB phase ultimately limits the amount of angular momentum that can be transfered onto the neutron star. This, in-turn, tunes the initial $P_{0}/\dot{P}_{0}$ fraction at birth (Figure 1, orange line). The dominant mechanism for energy loss (e.g., $n=3$ for pure dipole radiation) shapes the diagonal evolutionary path (Figure 1, green line).  The distribution of millisecond radio pulsars along the green diagonal line will be implicitly coupled with their Galactic birth rates ($R$). 

\subsection{Observed vs. projected \ppd distribution}

Synthetic millisecond pulsar populations that are constraint only by standard evolution are sampled from the parameter space ($D,\dot{M}, R$). In order to test whether the observed \ppd diversity can be reproduced, I integrate the consistent populations along with populations that show the 'most diverse' \ppd variations (Figure 2). The results can be summarized as follows:

\begin{figure}[t]
\includegraphics[angle=0,height=.25\textheight, trim=.3cm 1.5cm .1cm .75cm, clip=true ]{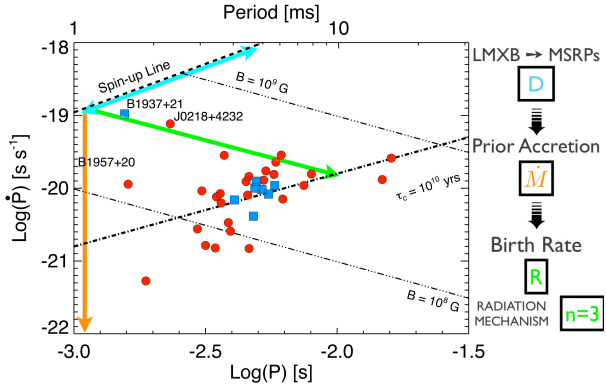} 	
\caption{Parameters that shape the \ppd demographics of millisecond pulsars.
}
\label{Fig:evol1}   
\end{figure}

\begin{itemize}\addtolength{\itemsep}{0.2\baselineskip}

\item  No physically motivated P$_{0}$ distribution has been able to produce the whole millisecond radio pulsar population consistently. No viable millisecond X-ray pulsar period distribution could subsequently produce the observed relative ratios of young/old pulsars with relatively high B fields. 

\item Observational selection effects cannot account for the rates at which higher magnetic field millisecond pulsars (e.g., PSR B1937+21) are observed. The fraction of observed young/old millisecond radio pulsars with high B fields is higher than what the standard model predicts by several orders of magnitude.

\item  Millisecond radio pulsars with characteristic ages $\tau_{c} > 10^{10}$~yr are born with P$_{0}\approx$ P and small spin-down rates. Figure 2 shows that the apparent enigma of millisecond pulsars with ages older than the age of the Galaxy is mainly a manifestation of very low accretion rates experienced during the late stages of their LMXB phase. 

\item  It is necessary to posit the existence of a separate class of progenitors, most likely with a different distribution of magnetic fields, accretion rates and equilibrium spin periods, presumably among the LMXBs that have not been revealed as millisecond X-ray pulsars. 

\item  A millisecond X-ray pulsar period distribution that has sharp multimodal features coupled with non-standard energy loss mechanisms may be able to reconcile for the joint \ppd distribution of millisecond pulsars.

\end{itemize}

\begin{figure}[t]
   \includegraphics[angle=90,height=.25\textheight]{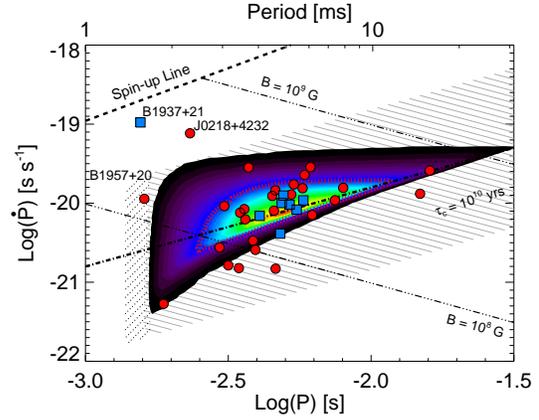} 
   \caption{Projected \ppd distribution of millisecond pulsars compared to observations.
}
\label{Fig:evol2}   
\end{figure}

\section{\label{sec:age} Age }

The standard method for estimating ages of pulsars is to infer an age from the rate of spin-down. The characteristic age ($\tau_{c}=P/2\dot{P}$) is only an approximation of the generic spin-down age and may give realistic estimates for normal pulsars for which the initial spin periods are likely to be much shorter than their observed periods (i.e., $P_{0} \ll P$). This method can fail, however, for pulsars with very short periods. For millisecond pulsars, the details of the spin-up process pose additional constraints on the period ($P$) and spin-down rates \pdp which may consequently affect the age estimate.

\subsection{A modified spin-down age for millisecond pulsars}

To estimate ages of millisecond pulsars, we propose the following method in which constraints arising from the recycling process and mass shedding are implemented:
\begin{eqnarray}
	\widetilde{\tau}\,(B > B_{c})&=&\frac{P}{(n-1)\;\dot{P}}\left[1-\left(\frac{\widetilde{\alpha}\, 
		\dot{P}^{3/7}}{P^{4/7}}\right)^{n-1}\right]\label{eq:mage1}			\\
	\widetilde{\tau}\, (B < B_{c})&=&\frac{P}{(n-1)\;\dot{P}}\left[1-\left(\frac{P_{sh}}{P}\right)^{n-1}\right]\label{eq:mage2}
\end{eqnarray}
where B$_{c}$ is the locus of points extending from the intersection of the diagonal spin-up line and the vertical mass shedding limit. P$_{sh}$ is the mass shedding limit which is expected to be between 0.85--1.32 ms depending on the equation of state of the NS matter \citep{Cook:94, Chakrabarty:03}.  Although $\widetilde{\alpha}$ inherently has numerous sources of uncertainty, the corresponding minimum post-accretion period that defines the spin-up line depends on the total accreted mass, but is insensitive to other parameters \citep[see,][]{Phinney:94}. Therefore, we parametrically adopt the re-normalized coefficient $\widetilde{\alpha}=2.6^{+0.7}_{-0.4} \times10^{6}{\bf s^{4/7}}$ and use it as a fiducial value.

While the modified spin-down age $\widetilde{\tau}$ gives a better estimate for the true age than $\tau_{c}$, it is still a strict upper limit (as is $\tau_{c}$ for MSPs). The projected underlying age distribution for standard evolution is reproduced in Figure 3. The conclusions are summarized as follows:

	\begin{figure}[]
	\includegraphics[width=2.3in,angle=90]{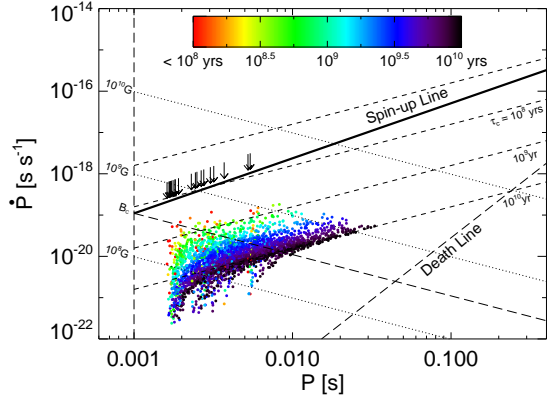}
	\caption{Expected true age distribution for millisecond pulsars. Color represents the true age $\tau_{t}$. Downward arrows ($\downarrow$) are neutron star spin frequencies measured in LMXBs which are used as progenitor seeds to reconstruct the synthetic population.
\label{fig:age}   
   }	
	\end{figure}

%\subsection{Age: Summary}

\begin{itemize}\addtolength{\itemsep}{0.2\baselineskip}

\item Characteristic age estimates are strict upper limits for millisecond pulsars.

\item A significant fraction of millisecond pulsars are born with apparent older ages.

\item Approximately 30\% of the population is expected to be born below the Hubble line ($\tau_{c}>10^{10}$yr). These millisecond pulsars must have experienced very low accretion rates during the LMXB phase.

\item There are two sources of age corruption: (1) {\it Age bias}, which is due to secular acceleration. Older millisecond pulsars may mimic younger ages because the measured spin-down rates ($\dot{P}$) may appear to spin-down faster due to this effect, significantly in some cases.  This bias can be corrected for millisecond pulsars with well determined distances and transverse velocities. (2) {\it Age contamination}, which is primarily driven by low accretion rates experienced during the LMXB phase. A significant fraction of the millisecond pulsar population is expected to mimic older ages due to this contamination. While a statistical assessment of the cumulative contribution of the ``age contamination'' (for the population) is possible \citep{Kiziltan:10}, there are no trivial methods which can be used to estimate the contribution to an individual source.

\item We estimate that $\sim$30\% of millisecond pulsars have ages that are overestimated by a factor of two or more. As a consequence, the NS-NS merger rates based on the characteristic ages of binary pulsars are mere underestimates.

\item While a braking index of n=3 is consistent with the millisecond pulsar \ppd demographics, it is challenging to disentangle the combined effects of a possible early phase during which energy was lost by more efficient processes such as gravitational wave radiation (n=5), and a subsequent secular evolution with torques less efficient than dipole radiation (n<3).
\end{itemize}

\section{\label{sec:mass} Mass }

In recent years, the number of pulsars with secure mass measurements has increased to a level which allows us to probe the underlying neutron star (NS) mass distribution in detail. Here, I briefly review the range of expected masses and how they compare to the mass distribution inferred from radio pulsar observations. A critical in-depth review with details on the numerical method can be found in \cite{Kiziltan:10g}.

\subsection{Birth, accreted and maximum mass}

Masses of NSs at birth are tuned by the intricate details of the astrophysical processes that drive core collapse and supernova explosions \citep{Timmes:96}. A careful analytical treatment of the uncertainties that affect the Chandrasekhar mass gives 1.08--1.57 \msun\,as a reasonable range for the birth mass \citep{Kiziltan:10g}. The expected amount of mass required to spin up pulsars to millisecond periods ($\Delta m_{acc}=0.10$--$0.20$ \msun) can be derived by angular momentum arguments \citep[for details see,][]{Kiziltan:10g}. For the upper mass limit, modern equations of states predict a range between 2.2--2.9 \msun. It is still unclear, however, whether very stiff equations of states (EOSs) that stably sustain cores up to the GR limit ($\sim3.2$ \msun) can exist. 

\subsection{Estimating the underlying mass distribution}

The methods used for NS mass measurements have intrinsically different systematics, and therefore require a more careful treatment when assessing the underlying mass distribution. There is no compelling argument to have a single coherent (unimodal) mass distribution for NSs that, we know, have  gone through dissimilar evolutionary histories and possibly different physical processes that lead to their production \citep{Podsiadlowski:04}. With an ever increasing number of pulsars with secure mass measurements, we have reached an adequate number by which we can now test whether the implied mass distributions are consistent with different sub-populations. 

\begin{figure}[]
\includegraphics[width=2.3in,angle=90, trim=.cm .cm .3cm .4cm, clip=true ]{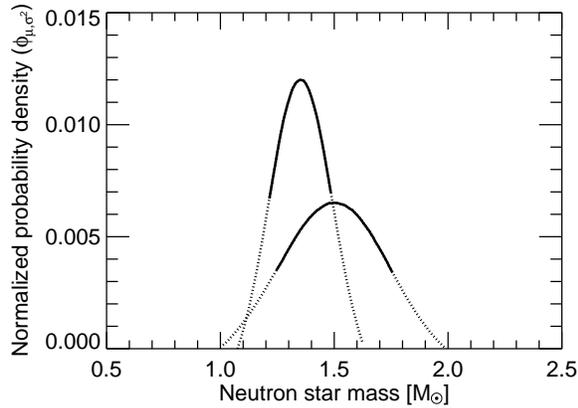} 	
\caption{Neutron star mass distribution. The underlying mass distribution of NSs show bi-modality with peaks at 1.35 \msun\,and 1.50 \msun\,for NSs in DNS and NS-WD systems respectively. The solid lines show the central 68\% probability range \citep[as in][]{Kiziltan:10g}.%The solid and dotted lines show the central 68\% and extended 95\% confidence ranges.
}
\label{Fig:post}   
\end{figure}

I take a comparative approach to analyze the available mass measurements both with conventional (maximum likelihood estimation) and modern (Bayesian) numerical methods. I use 

\bee
\lefteqn{\mathcal{L}(\mu,\sigma^{2};data)=} \nonumber\\
 & &\prod^{n}_{i=1}\left [2\pi (\sigma^{2}+S_{i}^{2})  \right]^{-1/2}\,e^{-\frac{(m_{i}-\mu)^{2}}{2(\sigma^{2}+S_{i}^{2})}}	
\label{Eq:like}
\eee
as the parametric likelihood and
\bee
\lefteqn{\mathcal{P}(\m_{0}|data) =}  \nonumber\\
 & &\int\int N(\m_{0};\mu,\sigma^{2})\, P(\mu,\sigma^{2}|data) \,d\mu \,d\sigma^{2}
\label{Eq:prob}
\eee
for the posterior inference, where for a mass distribution 
\bee
m_{i}=\mathcal{M}_{i}+w_{i}, i=1,...,n
\eee
$m$ is the pulsar mass estimate, $\mathcal{M}$ is the (unobserved) pulsar mass with an associated $w$ error.  For independent $\mathcal{M}_{i}$ and $w_{i}$, this yields
\bee
f(\mathcal{M},w)& = &N(\mathcal{M};\mu,\sigma^{2})\,N(w;0,S^{2}) \nonumber\\
&=&N(m;\mu,\sigma^{2}+S^{2})
\label{Eq:func}
\eee
where the associated errors are $w_{i}\sim N(0,S_{i}^{2})$.

Unlike conventional statistical methods, with a Bayesian approach it is possible to separately infer peaks, shapes and cutoff values of the distribution with independent reliability measures. This provides unique leverage to probe parameters which separately trace independent astrophysical and evolutionary processes. The inferred distribution is shown in Figure 4 and the ramifications can be summarized as follows:

\begin{itemize}\addtolength{\itemsep}{0.2\baselineskip}

\item Neutron star masses show peaks consistent with an initial mass function at 1.35$\pm0.14$ \msun\,and a recycled peak at 1.50$\pm0.25$ \msun. 

\item Considerable mass accretion ($\Delta m_{acc}\approx0.15$ \msun) has occurred during the spin-up phase.

\item A secure lower limit to the maximum mass of a NS exists at 2 \msun. This excludes the softer equations of states as viable configurations for matter at supra-nuclear densities.

\item The lack of truncation close to the maximum mass cutoff suggests that the maximum NS mass is set by evolutionary constraints rather than nuclear physics or general relativity, and the existence of rare super-massive NSs is possible. 

\end{itemize}

\begin{theacknowledgments}
I would like to thank the organizers of the ``Astrophysics of neutron stars---2010'' conference for the invitation. BK acknowledges NSF grant AST-0506453.
\end{theacknowledgments}

%%%%%%%%%%%%%%%%%%%%%%%%%%%%%%%%%%%%%%%
\bibliographystyle{aipproc}

\end{document}